# Enhanced Web User Interface Design Via Cross-Device Responsiveness Assessment Using An Improved HCI-INTEGRATED DL Schemes

Shrinivass Arunachalam Balasubramanian*

Senior Full Stack Engineer, Independent Researcher, United States, shrinivassab@gmail.com

User Interface (UI) optimization is essential in the digital era to enhance user satisfaction in web environments. Nevertheless, the existing UI optimization models had overlooked the Cross-Responsiveness (CR) assessment, affecting the user interaction efficiency. Consequently, this article proposes a dynamic web UI optimization through CR assessment using Finite Exponential Continuous State Machine (FECSM) and Quokka Nonlinear Difference Swarm Optimization Algorithm (QNDSOA). Initially, the design and user-interaction related information is collected as well as pre-processed for min-max normalization. Next, the Human-Computer Interaction (HCI)-based features are extracted, followed by user behaviour pattern grouping. Meanwhile, the CR assessment is done using FECSM. Then, the proposed Bidirectional Gated Luong and Mish Recurrent Unit (BiGLMRU) is used to classify the User eXperience (UX) change type, which is labelled based on the User Interface Change Prediction Index (UICPI). Lastly, a novel QNDSOA is utilized to optimize the UI design with an average fitness of 98.5632%. Feedback monitoring is done after optimal deployment.

## 1. INTRODUCTION

Currently, Websites have become significant platforms for UI, and the architecture of the websites enhances the UX in this user-oriented generation [1, 2]. Therefore, WD, a significant aspect of HCI, is utilized for improving the UX. This incorporates estimates of the User Behavior (UB) pattern, creating a user-friendly interface and testing the performance of the site [3, 4]. The UB patterns like screen size influence, and eye-tracking methodology are greatly helpful in improving the UX [5, 6]. By developing an age-friendly website, platforms like e-commerce improve the UX [7, 8]. However, traditional techniques did not concentrate on the CR assessment for web development [9, 10]. The proposed system's motivation is to develop a user-friendly website centered on the UB with an interface. Thus, a novel model for optimizing the website based on BiGLMRU is proposed in this paper.

### 1.1 Problem Statement

The prevailing works' limitations are given below,

- ❖ None of the works focused on CR assessment for web development.
- ❖ The UB patterns were not concentrated in [11], which mitigated the effectiveness of UI.
- ❖ The prevailing works designed the UI inefficiently, as the optimal UI design was not periodically updated.
- ❖ Existing works failed to focus on the improvement level of UI that affected the web development process.
- ❖ In [12], the support factors were not considered, which further reduced the effectiveness.

### 1.2 Objectives

The objectives of the proposed framework are defined below,

- ❖ The web is developed by considering the cross-responsiveness assessment using FECSM.
- ❖ To improve the effectiveness, the UB patterns are grouped by HDBSCAN.
- ❖ The optimal UI design is periodically updated by providing feedback to QNDSOA.

- ❖ By using BiGLMRU, the improvement level for the UI design is obtained.
- ❖ The support factors are considered by employing minimum JavaScript execution time, minimum error rate, and minimum memory usage as the fitness function in optimal UI design.

The remaining part is arranged as: in Section 2, the existing works are analyzed, in Section 3, the proposed methodology is explained, in Section 4, the results and discussion are given, and Section 5 concludes the paper with future scope.

## 2. LITERATURE SURVEY

Bakaev et al. [11] recognised the modeling of Visual Perception (VP) of the UI. Here, the VP was predicted by a Convolutional Neural Network (CNN). Yet, the change in responsiveness for different devices and screen sizes was not considered.

Todi et al. [12] assessed a Reinforcement Learning (RL) approach for adaptive UI. Firstly, both the positive and negative effects that impacted the UI were considered. Thus, adaptive UI adapted the webpage layouts and reorganized application menus. Nevertheless, the effectiveness of UI was reduced as the support factors based on optimal UI design were not considered.

Keselj et al. [13] examined the Deep Learning (DL) applications for the UI evaluation. Here, a CNN determined the effectiveness of UI based on the specifications, like UI design and layout. Yet, the user satisfaction was not achieved as the objective knowledge about UI was not practically implemented.

Muneer et al. [14] deployed a Meta-Model for supporting the Compatibility Testing (CT) of cross-browser web applications. Initially, for covering critical configurations, a checklist was initialized and translated into Interaction Flow Modeling Language (IFML). Next, the test cases generated by IFML addressed the compatibility issues.

Wang et al. [15] discovered a DL approach to assess the color quality interface in HCI interfaces. Firstly, the interface image features were extracted and modeled by a CNN. Yet, the UI design was still inefficient as the approach failed to update the user's immediate feedback.

## 3. PROPOSED METHODOLOGY FOR CROSS-RESPONSIVE WEB UI TUNING USING FECSM ANDBiGLMRU

The proposed work implements an intelligent framework for CR web UI optimization using FECSM and QNDSOA. In Figure 1, the proposed methodology's block diagram is presented.

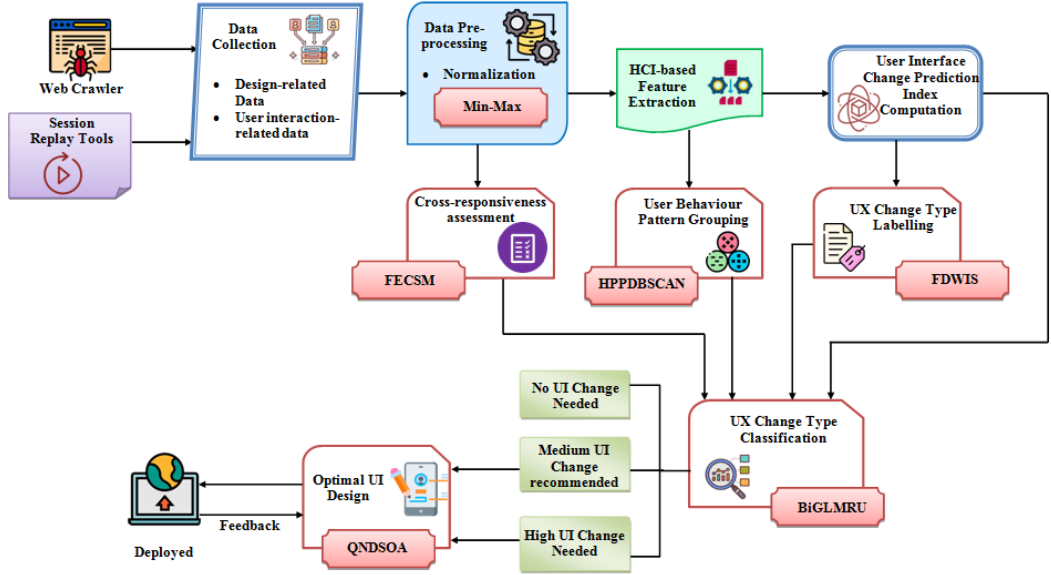

Figure 1: The structural design of the research approach

**3.1 Data collection**

Initially, the design-related data (layout, component arrangements, and responsiveness characteristics) and user interaction-related information (clicks, scrolls, and mouse movements) are collected by using the web crawler and session replay tools, respectively.

$$\partial_w = (\partial_1, \partial_2, \ldots \partial_W) \text{ Where, } w = 1 \, to \, W \tag{1}$$

Where, $W$ specifies the number of collected web data $\partial_w$.

**3.2 Pre-processing**

Next, $\partial_w$ is subjected to the pre-processing, which standardizes the collected data in the range of (0, 1) by employing min-max normalization.

$$\eta = \frac{\partial_w - \min(\partial_w)}{\max(\partial_w) - \min(\partial_w)} \tag{2}$$

Here, $\eta$ represents the pre-processed data.

### 3.3 HCI-based feature extraction

From $\eta$, the HCI features $(\gamma_i)$ like click patterns, scroll behaviour, mouse movement, and network condition are extracted, improving the model's performance.

### 3.4 User behaviour pattern grouping

Next, the $\gamma_i$ is fed into the proposed HPPDBSCAN approach, which groups the user behaviour pattern by considering the scroll rate and click depth. The conventional Hierarchical Density-Based Spatial Clustering of Applications with Noise (HDBSCAN) proficiently groups the data with varying density. However, the HDBSCAN is sensitive to the choice of the clustering parameters, like minimum cluster size $(\widetilde{X})$ and minimum samples $(Y)$. Therefore, the Persistence Probability Function (PPF) $(\wp)$ is used to determine the optimal parameters by analysing the density-based persistence of data across multiple scales.

$$\wp(\gamma_i) = \int_{S_k}^{S_K} |\gamma_i(S)| d''S \rightarrow (X, Y) \tag{3}$$

Where, $S_k$ indicate the density level $(S)$ where the cluster $\gamma_\infty$ appears, $S_K$ indicate the density level where the cluster $\gamma_\infty$ disappears, and $d''$ denotes the derivative parameter. For each point, the core point $(Cp)$ is computed based on the minimum number of neighbours, which is determined by $Y$. Similarly, the mutual reachability distance $(M_{dis})$ is estimated between the points to handle varying densities.

$$M_{dis}(\gamma_1, \gamma_2) = \max\{Cp(\gamma_1), Cp(\gamma_2), Z(\gamma_1, \gamma_2)\} \tag{4}$$

Where, $Z$ displays the direct distance value. By assigning $M_{dis}$ as the weight value of the edges, a complete mutual reachability graph $(U_{gr})$ is constructed. Subsequently, the minimum spanning tree $(T_{\min})$ is generated by connecting all points with the lowest $M_{dis}$ without creating any cycles. Next, the edges in the $T_{\min}$ are sorted by increasing the $M_{dis}$, and then the longest edges are gradually removed to create a hierarchical structure. Meanwhile, the tree pruning is done by applying $\widetilde{X}$ that evaluates the cluster's stability $(\lambda)$.

$$\lambda(T_{\min}) = \widetilde{X}(S_K - S_k) \tag{5}$$

Lastly, the data points are allocated to the clusters with the highest stability. Therefore, the user behaviour pattern grouped data is displayed as $(\phi_\nabla)$.

### 3.5 Cross responsiveness assessment

Contrarily, the CR assessment is done in $\eta$ using the proposed FECSM algorithm to model how interface responsiveness changes across devices over time. The Finite State Machine (FSM) significantly captures transitions in user experience due to changes in device configuration. Yet, the FSM struggled to handle continuous state and transition, affecting the model's flexibility. Therefore, the Exponential Continuous Coverage (ECC) function is utilized to handle transitions over changes.

Here, each state represents a responsive UI (mobile layout, tablet layout, and desktop layout).

$$St_v = (St_1, St_2, \ldots \ldots St_V) \text{ Where, } v = 1 \text{ to } V \tag{6}$$

Here, $V$ denotes the number of states $St_v$. Next, the inputs (user-triggering events) like mouse events, screen size changes, and touch gestures are defined as below,

$$Ip_w = \sum_{w=1}^{W} \{Ip_1, Ip_2, \ldots Ip_W\} \tag{7}$$

Here, $w = 1, 2, \ldots W$ indicates the number of inputs $Ip_w$. Also, the transitions are defined to reflect how the system moves from one state to another when an input is received. In the proposed work, the ECC function $(N)$ is used to ensure flexible transitions by continuously adapting to dynamic state changes.

$$N(\tau) = 1 - \exp^{-\Im \tau} \tag{8}$$

$$\tau \xrightarrow{\text{transition}} (St_1, St_v) \tag{9}$$

Here, $\Im$ specifies the controlling parameter and $\tau$ depicts the transitions. Subsequently, the start state and final state are also determined. Next, the user interactions are captured as a log file. Finally, the FSM trace logs are extracted to provide detailed insight into which transition caused friction and which device layout had higher task success. The CR assessed outcome is mentioned as $(H)$.

**3.6 User interface change prediction index computation**

Meanwhile, the UICPI $(\varphi)$ is calculated by considering the $\gamma_i$ to represent the necessity of UI modification based on user interaction deviation.

$$\varphi(\gamma_i) = v_1 \times E + v_2 \times T + v_3 \times D + v_4 \times C \tag{10}$$

Where, $(v_1, v_2, v_3, v_4)$ illustrates the weight values, $E$ denotes the error rate, $T$ depicts the task time, $D$ exhibits the drop-off rate, and $C$ represents the click confusion index.

**3.7 UX change type labeling**

The proposed Fuzzy Derivative Weighted Inference System (FDWIS) precisely labels the UX change type based on $\varphi$. The Fuzzy Inference System (FIS) offers high transparency. Yet, the FIS approach struggled to capture the small changes. Hence, the Derivative Weighted Average Function (DWAF) is employed in the defuzzification process to capture the small changes, improving the model's precision.

Initially, the fuzzification step converts the crisp values into fuzzy values $(\dddot{\varphi})$ (membership values) using a sigmoid membership function $(Q)$.

$$Q(\varphi) = \frac{1}{1 + \exp^{-G(\varphi - J)}} \rightarrow \dddot{\varphi} \tag{11}$$

Here, $G$ and $J$ denote the control parameter and center of the slope, respectively.

Here, the fuzzy if-then rules $(\Re_{rule})$ are created to categorize the UX change type based on the $\dddot{\varphi}$.

$$\Re_{rule} = \begin{cases} \text{If}(\dddot{\varphi} = 0.0 \text{ to } 0.3), & \text{then} \quad \text{Lw} \\ \text{If}(\dddot{\varphi} = 0.31 \text{ to } 0.6), & \text{then} \quad \text{Md} \\ \text{If}(\dddot{\varphi} > 0.6), & \text{then} \quad \text{Hw} \end{cases} \tag{12}$$

Next, the fuzzy rules are implemented in the fuzzified inputs to label the UX change type into low $(\text{Lw})$, medium $(\text{Md})$, and high $(\text{Hw})$.

Next, defuzzification is the task of converting the fuzzy outputs $(\varsigma)$ from the inference engine into a crisp value using DWAF. The DWAF prioritizes regions with sudden membership changes, causing improved precision.

$$\varphi = \frac{\sum_{c=1}^{C} O_c(\Re_{rule}) \cdot \left| \frac{\dot{b}\varsigma}{\dot{b}\dddot{\varphi}} \right| \cdot \varsigma}{\sum_{c=1}^{C} O_c(\Re_{rule}) \cdot \left| \frac{\dot{b}\varsigma}{\dot{b}\dddot{\varphi}} \right|} \tag{13}$$

Where, $O_c$ designates the firing strength of the $c^{th}$ rule, $\dot{b}$ depicts the partial derivative parameter, and $c = 1 \text{ to } C$ denotes the number of fuzzy rules.

**3.8 UX change type classification**

Here, $\phi_\nabla$, $H$, and $\varphi$ are inputted to the proposed BiGLMRU, which classifies the UI requirement into three categories like low UI changes needed, medium UI changes recommended, and high UI changes necessary based on the labelled data. The Bidirectional Gated Recurrent Unit (BiGRU) effectively captures the dynamic changes of the user interaction. Nevertheless, the BiGRU struggled to handle longer dependencies. Therefore, the Luong Attention (LA) function is used to hold long-term information. Likewise, the BiGRU had over-fitting issues. Hence, the Mish activation function is wielded to minimize the over-fitting issue by improving the gradient flow. In Figure 2, the proposed BiGLMRU's diagrammatic illustration is given.

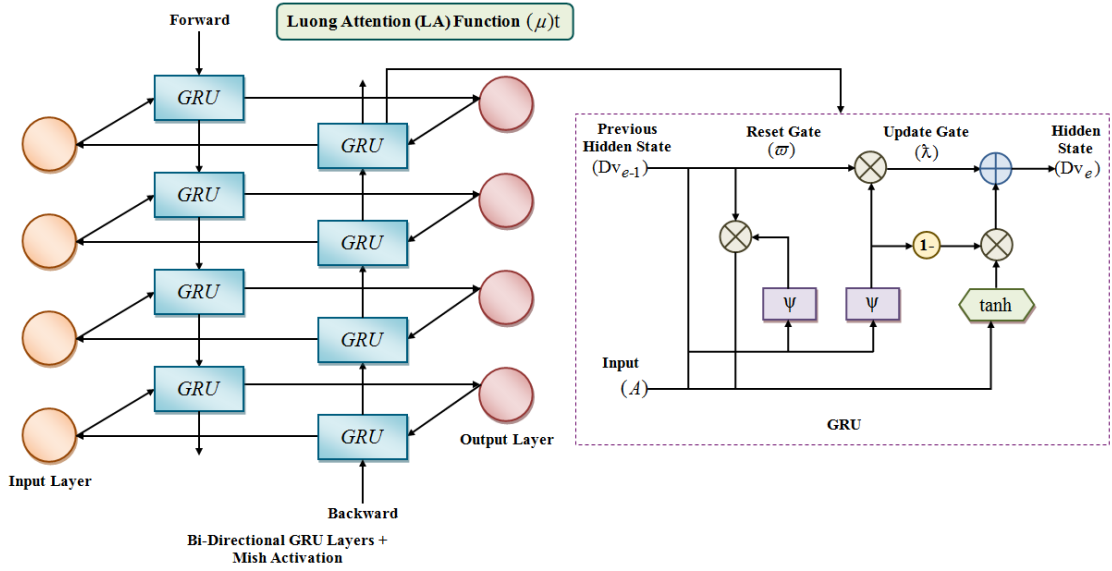

Figure 2: The pictorial depiction of the proposed BiGLMRU

The input layer $(A)$ holds the inputs as well as transmits them to the forward GRU layers.

$$A = (\phi_\nabla, H, \varphi) \tag{14}$$

The reset gate $(\varpi)$ aims to eradicate the less informative information in the previous hidden state $(Dv_{e-1})$. Likewise, the Mish activation function $(\psi)$ is employed to reduce the over-fitting issues due to its gradient stability.

$$\varpi = \psi \times ((A, Dv_{e-1}) \cdot Nu) + Xg \tag{15}$$

$$\psi(A) = A \cdot \tanh(\ln(1 + \exp^A)) \tag{16}$$

Here, $Nu$ and $Xg$ indicate the input's weight and bias, $\tanh$ exhibits the tangent function, and $\ln$ illustrates the logarithmic function. Likewise, the update gate $(\lambda)$ is used to include the relevant information in the present hidden state.

$$\hat{\lambda} = \psi \times ((A, Dv_{e-1}) \cdot Nu) + Xg \tag{17}$$

Also, it uses the LA function $(\mu)$ to advance the model's capability to capture the relevant information from the past sequences.

$$\mu = \tanh\left(Nu\left[\sum \chi \cdot \aleph; (A, Dv_{e-1})\right]\right) \quad (18)$$

Here, $\chi$ illustrates the softmax function and $\aleph$ depicts the probability score. Next, the candidate hidden state $\left(\tilde{D}v_e\right)$ is computed according to the $(A, Dv_{e-1})$, thereby holding long-term sequences.

$$\tilde{D}v_e = \tanh(\varpi \cdot (A, Dv_{e-1}) \cdot Nu) + Xg \quad (19)$$

Lastly, the hidden state $(Dv_e)$ is calculated by taking a weighted average of the previous and the candidate hidden state.

$$Dv_e = \mu \cdot \left((1 - \hat{\lambda}) \cdot Dv_{e-1} + \hat{\lambda} \cdot \tilde{D}v_e\right) \quad (20)$$

Besides, the output layer processes the input via the backward GRU layers. Here, the proposed BiGLMRU classifies the UI change requirement into low $(L_{UI})$, medium $(M_{UI})$, and high $(H_{UI})$ effectively.

$$\hat{v} = (L_{UI}, M_{UI}, H_{UI}) \quad (21)$$

Here, $\hat{v}$ establish the proposed BiGLMRU's outcome.

The proposed BiGLMRU's pseudo code is given below,

---

**Input:** $\phi_\nabla$, $H$ and $\varphi$
**Output:** UI change requirement classification
**Begin**
    **Initialize:** $\phi_\nabla$, $H$, $\varphi$ and $\psi$
    **For** 1 to each input do,
        **Determine** input layer
$$A = (\phi_\nabla, H, \varphi)$$
    *#GRU layers*
        **Execute** reset gate,
$$\varpi = \psi \times ((A, Dv_{e-1}) \cdot Nu) + Xg$$
        **Activate** Mish function,
$$\psi(A) = A \cdot \tanh(\ln(1 + \exp^A))$$
        **Perform** update gate $\hat{\lambda} = \psi \times ((A, Dv_{e-1}) \cdot Nu) + Xg$
        **Establish** LA function
$$\mu = \tanh\left(Nu\left[\sum \chi \cdot \aleph; (A, Dv_{e-1})\right]\right)$$
        **Compute** candidate hidden state
        **Estimate** hidden state
$$Dv_e = \mu \cdot \left((1 - \hat{\lambda}) \cdot Dv_{e-1} + \hat{\lambda} \cdot \tilde{D}v_e\right)$$

**Perform** output layer $\hat{v} = (L_{UI}, M_{UI}, H_{UI})$

  **End For**

**Return** $\hat{v}$

**End**

### 3.9 Optimal UI design

The proposed QNDSOA is established regarding the requirement of medium $M_{UI}$ and high UI changes $H_{UI}$ to optimize the UI design. The Quokka Swarm Optimization Algorithm (QSOA) is highly adaptable to adjust the parameters, like acceleration coefficients. But, the QSOA struggled to determine the position of the member across the population. Thus, the Nonlinear Difference Function (NDF) is used to reflect the diversity of the position across the population.

Initially, the population members are initialized in the search area. Here, the inputs like font size, theme mode, letter spacing, and text alignment are considered as the quokka (member).

$$L_x = \{L_1, L_2, \ldots \ldots L_X\} \text{ Where, } x = 1 \, to \, X \tag{22}$$

Here, $X$ demonstrates the number of population members $L_x$. In the proposed work, the minimum JavaScript execution time, minimum error rate, and minimum memory usage are considered as the fitness values to select the best leader $\left(L_x^{Best}\right)$. Then, the member's location and drought $(Dh)$ are updated regarding the $L_x^{Best}$. The proposed work introduces the NDF $(\alpha)$ to cover the diversity of the position across the population.

$$Dh^{new} = \frac{(Tm + hm)}{(0.8 + Dh)} + \iota \cdot \alpha \tag{23}$$

$$\alpha = \exp^{-O\hbar}\left(L_x^{Best} - L_x\right) \tag{24}$$

$$L_x^{new} = L_x + Dh^{new} * \sigma \tag{25}$$

Where, $Dh^{new}$ depicts the updated drought, $L_x^{new}$ demonstrates the updated member's position, $Tm$ indicates the temperature (balancing parameter), $hm$ illustrates the humidity (exploration force), $\sigma$ denotes the nitrogen ratio (solution quality), $\iota$ implies the weight between the leader and members, $O$ exhibits the adaptive parameter, $\hbar$ specifies the time bound, and $\alpha$ indicates the differences of position between the leader and quokka. Next, fitness is updated iteratively until it converges. The QNDSOA's pseudo code is given below,

**Input:** Web components
**Output:** Optimal web UI $(O_{UI})$

**Begin**

    **Initialize** $L_x, L_x^{Best}, Dh$ and $\sigma$

    **For** 1 to each member do,

        **Perform** population initialization $L_x = \{L_1, L_2, \ldots\ldots L_X\}$

        Select leader $L_x^{Best}$ via fitness value

        **Update** drought,

$$Dh^{new} = \frac{(Tm + hm)}{(0.8 + Dh)} + \iota \cdot \alpha$$

        **Apply** NDF $\alpha = \exp^{-O\hbar}\left(L_x^{Best} - L_x\right)$

        **Update** member's location

$$L_x^{new} = L_x + Dh^{new} * \sigma$$

        **Repeat** until converge

    **End For**

**Return** $O_{UI}$

**End**

The proposed QNDSOA optimizes UI layout by balancing responsiveness across devices. Once the optimized UI design is generated, it is deployed into the live web environment. User interaction with the new UI is continuously monitored, providing feedback for fine-tuning the model.

## 4. RESULTS AND DISCUSSION

The experimental investigation is done to validate the performance of the proposed work, which is implemented in the PYTHON platform.

### 4.1 Dataset description

The design-related information and user interaction-related information are gathered in real-time to evaluate the proposed approach using the web crawler and session replay tools. From the whole data, 80% as well as 20% of the data are allocated for training along with testing.

### 4.2 Performance assessment of the research methodology

The proposed work's performance is appraised with numerous prevailing algorithms.

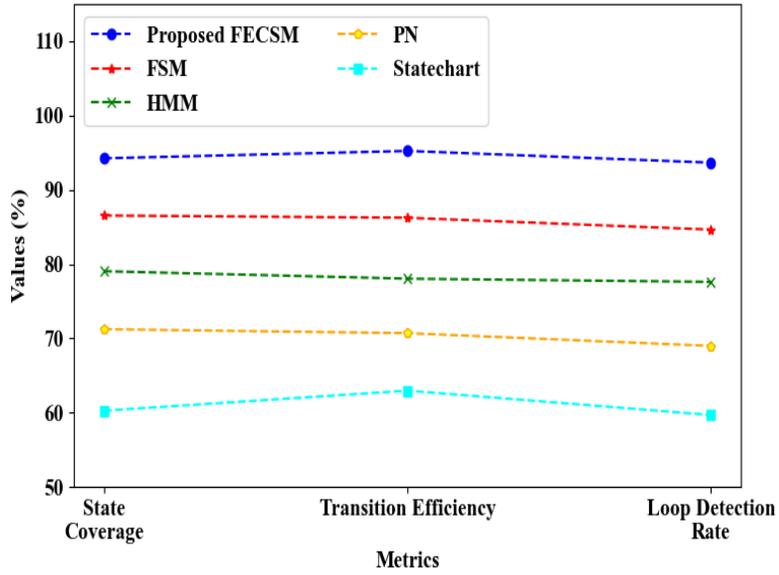

Figure 3: Empirical analysis for CR assessment

Table 1: Performance validation of the proposed FECSM

| Algorithm | State Coverage (%) | Transition efficiency (%) | Loop detection rate (%) |
|---|---|---|---|
| Proposed FECSM | 94.2356 | 95.2312 | 93.6532 |
| FSM | 86.5402 | 86.2356 | 84.6375 |
| HMM | 79.0364 | 78.0326 | 77.6023 |
| PN | 71.2341 | 70.6982 | 68.9782 |
| State chart | 60.2584 | 62.9584 | 59.6803 |

The proposed FECSM's performance is weighed against the prevailing FSM, Hidden Markov Model (HMM), Petri Net (PN), and state chart in Figure 3 and Table 1. The proposed FECSM attained state coverage, transition efficiency, and loop detection rate of 94.2356%, 95.2312%, and 93.6532%, respectively. Contrarily, the traditional FSM had state coverage, transition efficiency, and loop detection rate of 86.5402%, 86.2356%, and 84.6375%, respectively, showing limited adaptability. Here, the presence of ECC aided in improving the proposed work's performance in CR assessment.

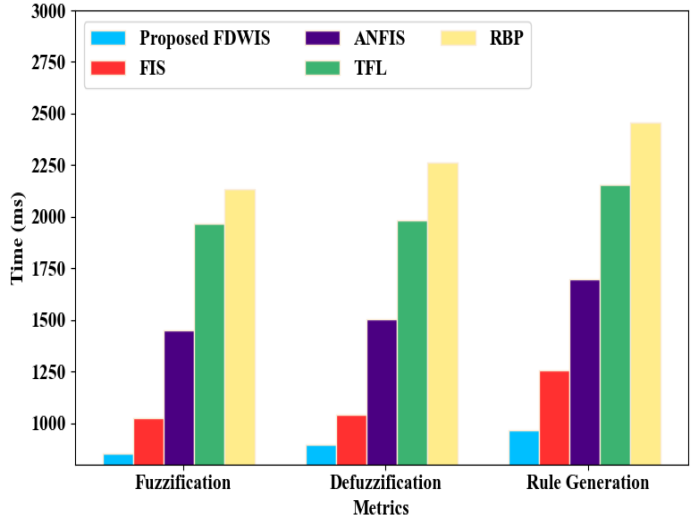

Figure 4: Performance evaluation for UX change labeling

Table 2: Numerical investigation of the proposed FECSM

| Techniques | Fuzzification time (ms) | Defuzzification time (ms) | Rule generation time (ms) |
|---|---|---|---|
| Proposed FDWIS | 855 | 897 | 963 |
| FIS | 1024 | 1042 | 1255 |
| ANFIS | 1450 | 1501 | 1698 |
| TFL | 1964 | 1985 | 2157 |
| RBP | 2135 | 2264 | 2455 |

In Figure 4 and Table 2, the proposed FDWIS's performance is appraised with prevailing FIS, Adaptive-Neuro FIS (ANFIS), Trapezoidal Fuzzy Logic (TFL), along with Rule-Based Prediction (RBP) to exhibit the model's supremacy in UX change labeling. The presence of DWAF-based defuzzification upgraded the efficacy of the labeling. The proposed FDWIS had fuzzification time, defuzzification time, along with rule generation time of 855ms, 897ms, and 963ms, respectively. But, the traditional techniques had maximum time consumption. Therefore, the FDWIS's dominance was evidenced.

Table 3: Comparative assessment for UX change type classification

| Methods | Accuracy (%) | Precision (%) | Recall (%) | F-Measure (%) | Sensitivity (%) | Specificity (%) |
|---|---|---|---|---|---|---|
| Proposed BiGLMRU | 99.2315 | 98.0745 | 98.1265 | 98.1005 | 98.1265 | 98.0745 |
| BiGRU | 93.5489 | 90.4568 | 91.6544 | 91.0556 | 91.6544 | 90.4568 |
| LSTM | 88.9746 | 87.4571 | 88.0267 | 87.7419 | 88.0267 | 87.4571 |
| RNN | 84.3922 | 81.9655 | 80.9472 | 81.4563 | 80.9472 | 81.9655 |
| DLNN | 77.3586 | 76.3204 | 77.5543 | 76.9373 | 77.5543 | 76.3204 |

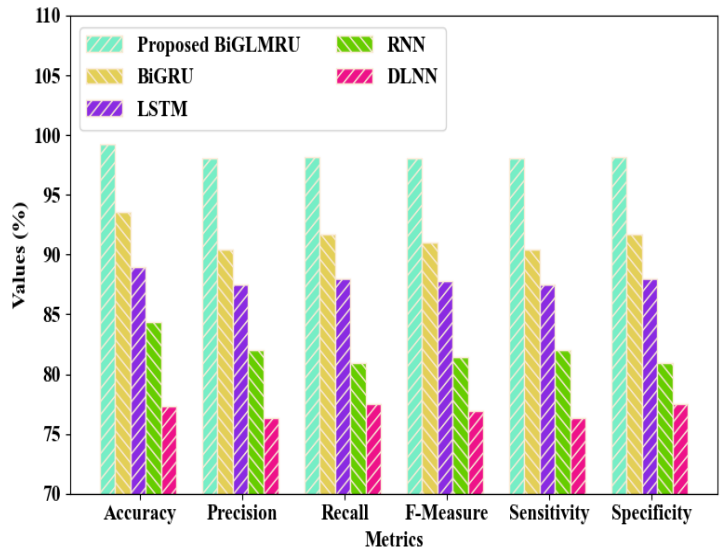

Figure 5: Performance assessment of the proposed BiGLMRU

In Figure 5 and Table 3, the proposed BiGLMRU's performance is appraised with the prevailing BiGRU, Long Short Term Memory (LSTM), Recurrent Neural Network (RNN), and Deep Learning Neural Network (DLNN). Regarding accuracy, precision, recall, f-measure, sensitivity, along with specificity, the BiGLMRU attained 99.2315%, 98.0745%, 98.1265%, 98.1005%, 98.1265%, and 98.0745%; while, the prevailing techniques attained 86.0685%, 84.0499%, 84.5456%, 84.2977%, 84.5456%, and 84.0499%. The existing works obtained poor classification performance. But, the BiGLMRU utilized Mish activation function for mitigating the over-fitting issues, enhancing the model's superiority.

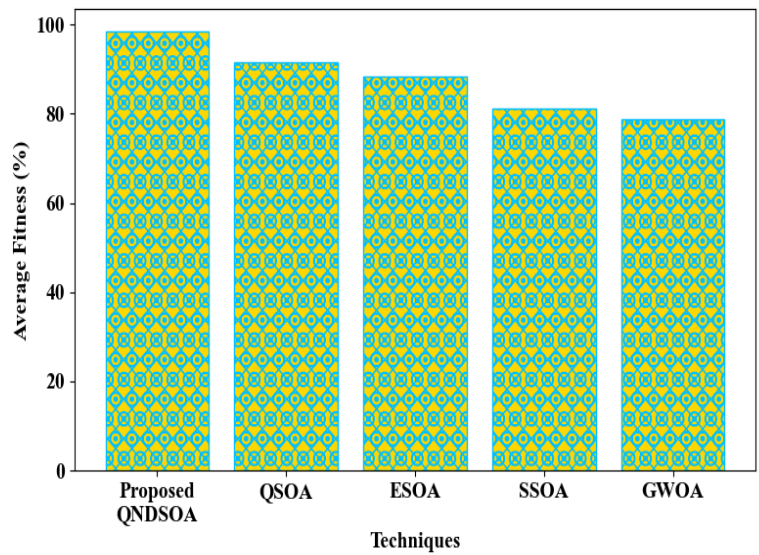

Figure 6: Average fitness analysis

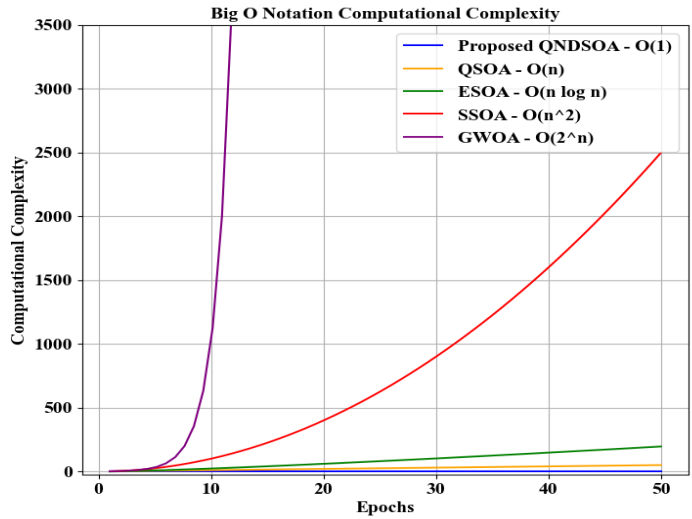

Figure 7: Computational complexity analysis for UI optimization

The proposed QNDSOA's performance is weighed against the prevailing QSOA, Egret Swarm Optimization Algorithm (ESOA), Salp Swarm Optimization Algorithm (SSOA), along with Grey Wolf Optimization Algorithm (GWOA) in Figures 6 and 7. The proposed QNDSOA achieved an average fitness of 98.5632%, whereas the traditional GWOA had 78.6594%. Further, the proposed work had limited complexity regarding varying epochs due to the utilization of NDF-based position updation. Thus, the QNDSOA had better performance.

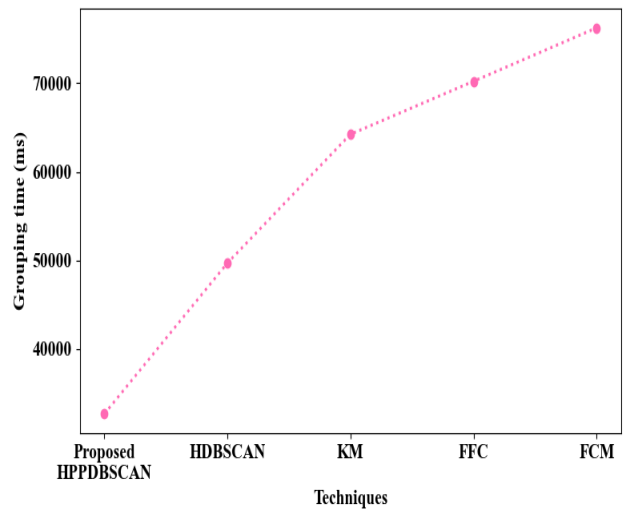

Figure 8: Performance assessment for user behavior pattern grouping

In Figure 8, regarding grouping time, the proposed HPPDBSCAN's performance is weighed against the prevailing HDBSCAN, K-Means (KM), Farthest First Clustering (FFC), and Fuzzy C-Means (FCM). The proposed HPPDBSCAN took 32654ms to complete grouping, whereas the existing HDBSCAN obtained a grouping time of 49687ms. Therefore, the proposed work had low time complexity due to the effectual parameter selection.

### 4.3 Comparative validation of the proposed work

The research methodology's comparative analysis is done to exhibit the model's prominence.

Table 4: Comparative validation

| Author's name | Target area | Methods | Merits | Challenges |
|---|---|---|---|---|
| Ma [16] | Computer web interface optimization | BPNN | Faster page loading | Script execution delays |
| Wang [17] | Intelligent layout adaptation in web page design | NCMF | Higher user engagement | Dynamic Content Instability |
| Kikuchi et al. [18] | Enhanced web page layout optimization | Optimization-based hierarchical layout mode | Improved responsive design | Layout shift issues |
| Martin et al. [19] | Personalized web UI adaptation | Situation adaptation-aware scheme | Better flexibility and adaptability | Over-responsive elements |
| Xu & Wang [20] | Interactive website search interface design | Concave-convex texture mapping algorithm | Enhanced web accessibility | Less adaptability |
| Proposed work | Enhanced web UI design via CR assessment using an advanced HCI | FECSM and QNDSOA | Improved cross-compatibility and adaptive layout design | The proposed work heavily relied on UI optimization rather than interpretability |

In Table 4, the proposed work's performance is compared with several associated studies. The proposed FECSM and QNDSOA algorithms aided in improving the user experience of the web environment through CR assessment. Similarly, to optimize the computer web interface, the existing works utilized Back Propagation Neural Network (BPNN) (Ma, 2022) and Non-negative Convolutional Matrix Factorization (NCMF) (Wang, 2022). Nevertheless, the existing work had adaptability issues and computational overhead. Thus, the proposed work achieved adaptive layout design with less complexity.

### 5 CONCLUSION

Here, this article proposed an enhanced web UI design through CR assessment using an improved HCI-integrated FECSM and QNDSOA approaches. The proposed FECSM provided detailed insight into the transitions across different screen sizes with a state coverage of 94.2356%. Similarly, a novel QNDSOA significantly optimized the web UI with an average fitness of 98.5632%. Besides, the constant feedback monitoring was enabled to ensure the model's trustworthiness. Nevertheless, the proposed work primarily focused on optimizing the UI design rather than interpretation.

*Future scope:* Thus, this work will focus on considering explainable AI and cognitive load factors in the future to improve the reliability and trust of the UI-enhancement process.